\documentclass[a4paper]{aa}
\usepackage{natbib}
\usepackage{graphicx}
\usepackage{color}

\title{Galactic distributions of carbon- and oxygen-rich AGB stars
revealed by the AKARI mid-infrared all-sky survey}
\author{
Daisuke Ishihara$^{1}$,
Hidehiro Kaneda$^{1}$,
Takashi Onaka$^{2}$,
Yoshifusa Ita$^{3}$,
Mikako Matsuura$^{4, 5}$,\\
\and
Noriyuki Matsunaga$^{6}$}
\institute{
Department of Physics,
Nagoya University, Furo-cho, Chikusa-ku, Nagoya, Aichi, 464-8602, Japan
\and
Department of Astronomy, Graduate School of Science,
University of Tokyo, 7-3-1 Hongo, Bunkyo-ku, Tokyo, 113-0033, Japan
\and
Astronomical Institute, Tohoku University,
6-3 Aramaki, Aoba-ku Sendai, 980-8578, Japan
\and
UCL-Institute of Origins, Department of Physics and Astronomy,
University College London, Gower Street, London
WC1E 6BT, UK
\and
UCL-Institute of Origins, Mullard Space Science Laboratory,
University College London, Holmbury
St. Mary, Dorking, Surrey RH5 6NT, UK
\and
Kiso Observatory, Institute of Astronomy,
University of Tokyo, 10762-30,
Mitake, Kiso, Nagano 397-0101, Japan
}

\abstract
{The environmental conditions
for asympotic giant branch (AGB) stars
to reach the carbon-rich (C-rich) phase
are important 
to understand
the evolutionary process of AGB stars.
The difference between the spatial distributions
of C-rich and oxygen-rich (O-rich) AGB stars
is essential for 
the study of the Galactic structure
and the chemical evolution of the interstellar medium (ISM).
}
{
We quantitatively investigate
the spatial distributions of
C-rich and O-rich AGB stars in our Galaxy.
We discuss the difference between them
and its origin.
}
{
We classify a large number of 
AGB stars newly detected 
by the AKARI mid-infrared all-sky survey.
In the color-color diagrams,
we define their occupation zones
based on
the locations of known objects.
We then obtain the spatial distributions of
C-rich and O-rich AGB stars,
assuming that they have the same luminosity
for a given mass-loss rate.
}
{
We find that
O-rich AGB stars are concentrated toward the Galactic center
and that the density decreases with Galactocentric distance,
whereas C-rich AGB stars show a relatively uniform distribution
within about 8\,kpc of Sun.
}
{
Our result confirms the trends
reported in previous studies
and extends them to a Galactic scale.
We discuss
the relations between our result,
the Galactic metallicity gradient,
and the chemical evolution of the ISM in our Galaxy.
}
\authorrunning{Ishihara, D. et al.}
\titlerunning{Galactic distribution of AKARI sources}

\bibpunct{(}{)}{;}{a}{}{,}

\begin{document}
\maketitle

\section{Introduction} \label{intro}

Asymptotic giant branch (AGB) stars
are low- to intermediate-mass (1--8\,M$_{\odot}$) stars
at their final evolutionary stages
that undergo mass-loss \citep[e.g.][]{ARAA}.
Their ages range from
0.1 to several Gyr,
depending on their initial masses.
The spatial distribution of AGB stars
roughly traces past star-formation activities.

The AGB stars are divided into three classes
based on
their carbon-to-oxygen 
elemental
abundance ratios:
carbon-rich AGB stars (C-rich; C/O$>$1),
oxygen-rich AGB stars (O-rich; C/O$<$1),
and S stars (C/O$\sim$1).
Many AGB stars in the Milky Way
are O-rich
\citep[e.g.][]{ARAA}
and show silicate-dominated circumstellar features.
In contrast,
some fraction of
intermediate-mass AGB stars (1.1--4\,M$_{\odot}$)
become C-rich,
displaying
the features of both carbon-bearing molecules
(e.g. C$_2$, CN, HCN, C$_2$H$_2$, and CO)
and carbonaceous dust (e.g. amorphous carbon and graphite).
This chemical change is triggered
when carbon atoms, which are synthesized in the stellar core,
are dredged up to the surface
by the deep convective envelope.

The difference in the spatial distributions
between O-rich and C-rich AGB stars,
i.e., the variation in the ratio of
C-rich AGB stars to O-rich AGB stars
(C/M ratio) from place to place,
has long been studied
\citep{Blanco,Westerlund}.
It is known that
O-rich AGB stars concentrate
toward the Galactic center (G.C.),
while C-rich AGB stars 
are distributed more uniformly around Sun
\citep[e.g.][]{Thronson,Claussen,Jura,Noguchi}.
For example,
\citet{IRTS2} investigated
the spectra of 689 AGB stars
acquired in the four radial directions
($l$=$-10^\circ$, $50^\circ$, $130^\circ$, and $170^\circ$)
with the Near-Infrared Spectrometer \citep[NIRS;][]{NIRS}
on-board the InfraRed Telescope in Space \citep[IRTS;][]{IRTS},
and found that
O-rich AGB stars outnumber C-rich ones
in the inner Galaxy ($<$8\,kpc), whereas
the situation is the reverse in the outer Galaxy.
However,
most past studies surveyed limited
regions of the solar neighborhood
because of their sensitivity constraints.
A comprehensive study of O-rich and C-rich AGB stars
covering the entire Milky Way
and a quantitative discussion of their
dependence on environment (e.g. metallicity) 
has not been performed yet.

The galactic AGB stars are believed to be a significant 
contributor to the interstellar medium (ISM).
Their distribution and amount of mass loss
are interesting in view of
the chemical evolution of the ISM in our Galaxy and
the life cycle of heavy elements in the universe.
%
%
Recently, supernova remnant (SNR)
is focused as the candidate to fill the missing budget
in the supply of the interstellar dust
\citep[e.g.][]{CasA,HoGyu,Tycho}.
In particular,
many discussions have focused on the supply of dust
from the remnants of core-collapse (Type II) supernovae
since the discovery of dust in the early universe \citep{Bertoldi}.
However,
the supply of carbonaceous grains should be discussed
independently of the supply of the total amount of dust grains
dominated by silicates
to clarify the whole picture of the life cycle of ISM
and explore clues to the origin of life.
The gas and dust budgets and their impact on the ISM evolution
were quantitatively investigated by \citet{Mikako}
using sensitive Spitzer images
of the Large Magellanic Cloud (LMC),
and the total mass loss amount from C-rich AGB stars revised.

From 2006 to 2007,
AKARI \citep{AKARI} surveyed the entire sky
with six infrared photometric bands between 9--160\,$\mu$m.
The mid-infrared (MIR) part of the all-sky survey was performed
with two broad bands centered at 9 and 18\,$\mu$m \citep{ScanOpe} 
using one of the on-board instruments, the InfraRed Camera \citep[IRC;][]{IRC}.
The detection limits (5$\sigma$) for point sources per scan are
50 and 90\,mJy for the 9 and 18\,$\mu$m bands, respectively,
with spatial resolutions of about 5$''$,
thus surpassing
the IRAS survey in the 12 and 25\,$\mu$m bands
by an order of magnitude
in both sensitivity and spatial resolution.
More than 90\% of the entire sky was observed at both bands
during the lifetime of the liquid helium cryogen.
In total, 870,973 sources
(844,649 in the 9\,$\mu$m band and 194,551 in the 18\,$\mu$m band)
were extracted with the uniform threshold
and compiled as the first point-source catalogue \citep[PSC;][]{MirCat}.
%
\citet{Yita} reported
that the MIR catalogue contains at least
90,000 AGB stars.

In this paper,
we investigate
the spatial distributions of
C-rich and O-rich AGB stars across the Galaxy.
We describe the data analysis applied 
to select C-rich and O-rich AGB samples
from the AKARI/MIR all-sky survey PSC (hereafter AKARI/MIR PSC)
and the reliability of the samples in Sect.~2.
The spatial distributions of and any differences
between C-rich and O-rich AGB stars in our Galaxy
are presented in Sect.~3.
Section~4 discusses
the origin of the difference in
the distributions of C-rich and O-rich AGB stars,
compares our results with related and extragalactic results,
and describes the effects on the chemical evolution of the ISM in our Galaxy.
A summary is given in Sect.~5.

\section{Data analysis} \label{obs}

Source classifications
based on the loci of sources in color-color diagrams
were often applied to previous sets of infrared survey data for
IRAS \citep[e.g.][]{WalkerA,WalkerB,IRAScc},
MSX \citep[e.g.][]{Egan, Lumsden}, and Spitzer \citep[e.g.][]{Spitzercc}.
This type of method basically utilizes the characteristics of
the spectral energy distributions (SEDs) of the detected sources.
It is an effective and unique approach
in classifying a large amount of unidentified objects
with only photometric data,
though contamination is inevitable.
We apply this method
to the AKARI/MIR PSC,
which contains many unidentified objects,
and classify all the detected objects,
selecting C-rich and O-rich AGB stars among them.

We initially cross-identify all
9\,$\mu$m and 18\,$\mu$m sources
in the AKARI/MIR PSC $\beta$1 data set \citep{MirCat}
with the 2MASS Point Source Catalogue version 7 \citep{2MASS}
based
simply on given celestial coordinates.
The search radius of 3$''$ was optimized
for the cross-identification 
between these two catalogues \citep{MirCat}.
In total, 614,204 AKARI/MIR
sources are cross-identified with the 2MASS catalogue.
We also search for the identification of the AKARI/MIR sources
in the SIMBAD database. The search radius is 10$''$.
In total, 331,764 AKARI/MIR sources 
have plausible identifications.
In other words, more than 60\% of the 
AKARI/MIR sources are new 
compared to the SIMBAD database.

\begin{figure*}[h]
(a) [J]--[K] vs. [K]--[9]\hspace{2.9cm}
(b) [K]--[9] vs. [9]--[18]\hspace{2.9cm}
(c) [J]--[K] vs. [9]--[18]\\
\includegraphics[width=6cm]{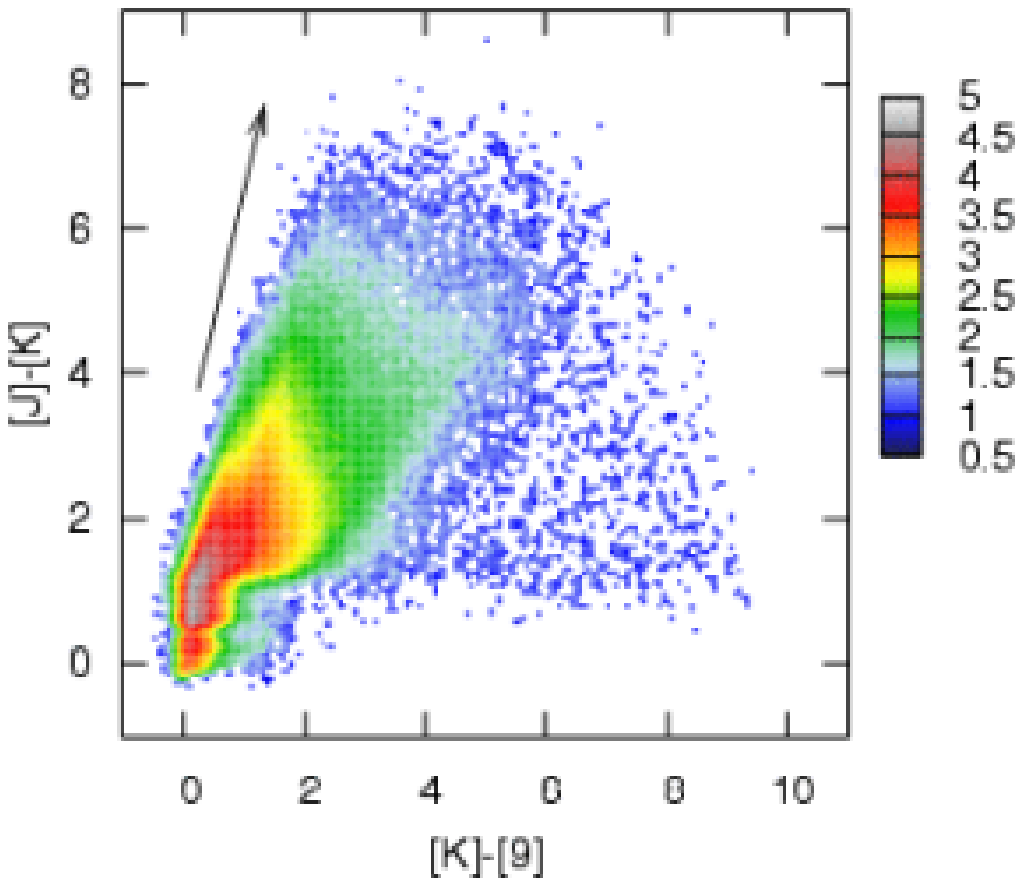}
\includegraphics[width=6cm]{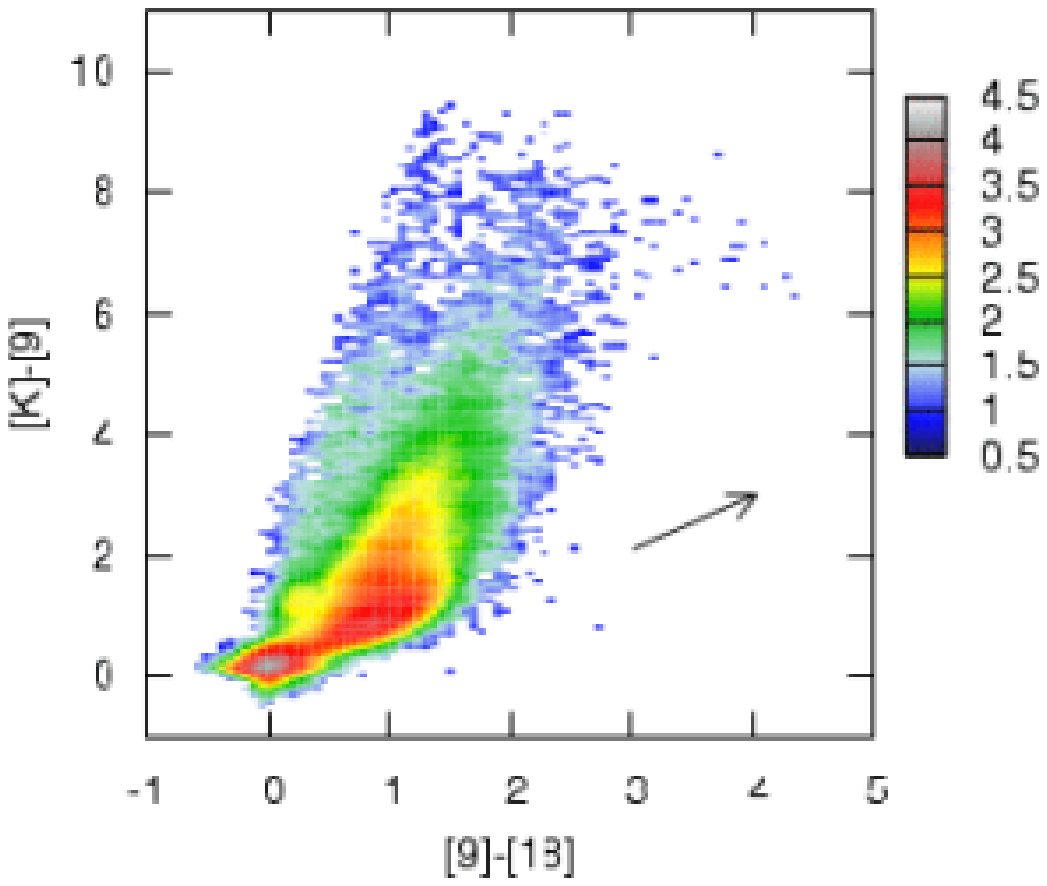}
\includegraphics[width=6cm]{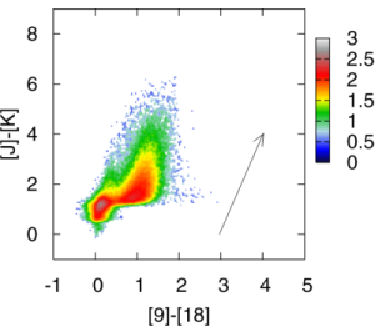}\\
\caption{Color-color diagrams in density maps 
for the AKARI/MIR sources that have 2MASS measurements:
(a) [J]--[K] versus [K]--[9],
(b) [K]--[9] versus [9]--[18], and
(c) [J]--[K] versus [9]--[18] color-color diagrams.
The arrow on each panel shows
the interstellar extinction vector for $A_{\rm V}=20$ mag,
by using the \citet{Av} Milky Way model of $R_{\rm V}=3.1$.
The color scales of the maps are logarithmic in number per unit color space.
(The color versions of this and subsequent figures
are available in the online journal.)
\label{fig:cc_density}}
\end{figure*}

\begin{figure*}[h]
\center
\includegraphics[width=18cm]{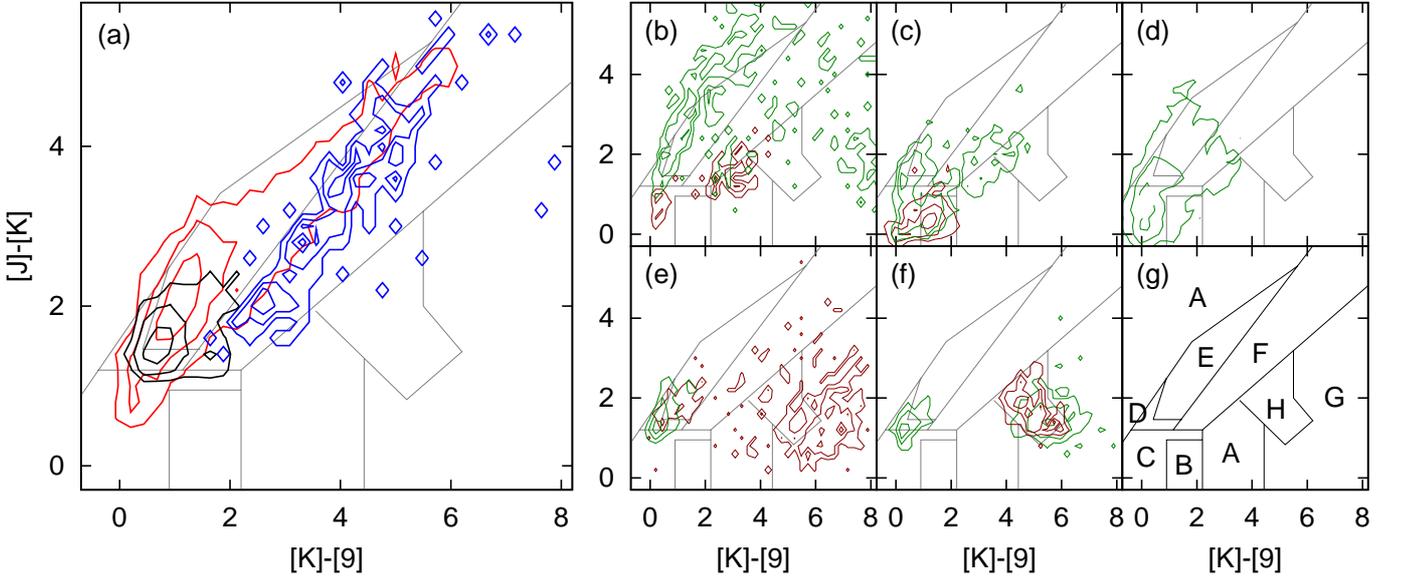}
\caption{Distributions of the known members
(identified in the SIMBAD database) 
shown in contours
on the [J]--[K] versus [K]--[9] color-color diagram :
(a) C-rich AGBs (red; group E), AGBs (black) and OH/IR stars (blue) (O-rich AGBs; group F);
(b) T Tauri (brown) and YSOs (green) (group A);
(c) Be stars (including Wolf-Rayet stars; brown) and
emission-line stars (green) (group B);
(d) normal stars including O--M giants and dwarfs (green; group C);
(e) S-stars (green; group D), and post-AGBs and planetary nebulae (brown; group G); and
(f) normal and starburst galaxies (green) and Seyferts (brown) (group H).
The occupation zones for the groups A--H are summarized in (g).
The contours show 20$\%$, 50$\%$, and 80$\%$ of the peak value.
The SIMBAD definitions of the members of
the groups A--H are summarized in Table~\ref{tbl:cc}.
\label{fig:ccds1}}
\end{figure*}

\begin{figure*}[h]
\center
\includegraphics[width=18cm]{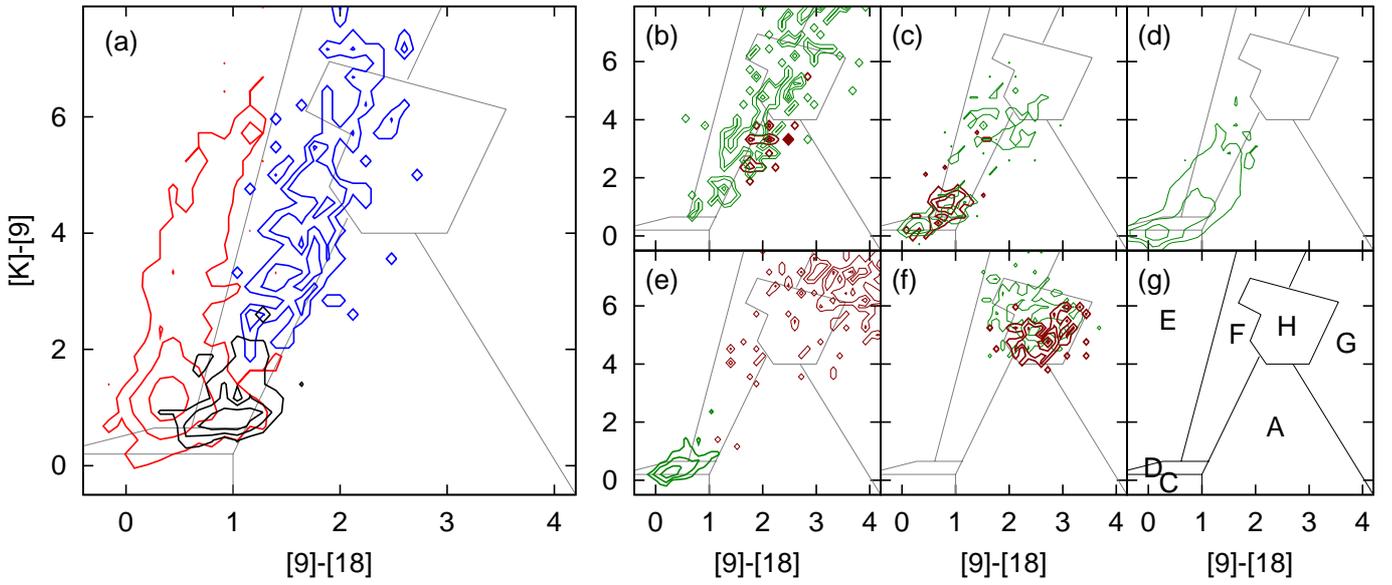}
\caption{Same as Fig.~\ref{fig:ccds1}, but 
the data are instead presented
on the [K]$-$[9] vs. [9]$-$[18] color-color diagram.
The occupation zone for emission-line stars (B)
is not defined on this diagram because
they cannot be clearly separated from the other objects (panel c).
\label{fig:ccds2}}
\end{figure*}

\begin{figure*}[h]
\center
\includegraphics[width=18cm]{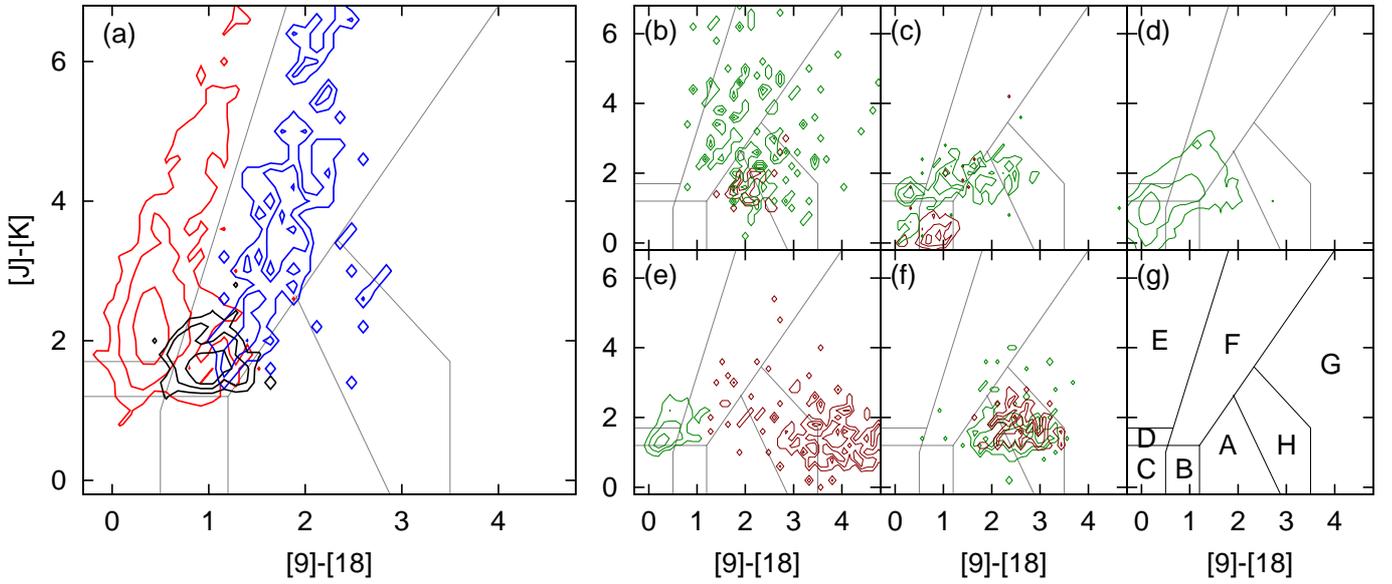}
\caption{Same as Fig.~\ref{fig:ccds1},
but shown on the [J]$-$[K] vs. [9]$-$[18] color-color diagram.
\label{fig:ccds3}}
\end{figure*}

All the cross-identified sources
are plotted on color-color diagrams.
Figures~\ref{fig:cc_density}a, \ref{fig:cc_density}b,
and \ref{fig:cc_density}c show the
color-color diagrams of
[J]--[K] versus (vs.) [K]--[9],
[K]--[9] vs. [9]--[18], and
[J]--[K] vs. [9]--[18], respectively.
Here, [J], [K], [9], and [18] represent Vega magnitudes in the
2MASS J, 2MASS K$_{\rm s}$, AKARI 9\,$\mu$m, and AKARI 18\,$\mu$m bands,
respectively.
The zero point magnitudes for the 9\,$\mu$m and 18\,$\mu$m bands are
56.3 and 12.0\,Jy, respectively \citep{IRC}.

We then define eight groups (A--H)
and assign already known objects to the groups
following the identification of the SIMBAD database
as summarized in Table \ref{tbl:cc}.
Figures~\ref{fig:ccds1}, \ref{fig:ccds2}, and \ref{fig:ccds3}
show the distributions of the known objects
on the same color-color diagrams 
as Figs.~\ref{fig:cc_density}a, \ref{fig:cc_density}b,
and \ref{fig:cc_density}c,
respectively.
In all three diagrams,
a large fraction of objects are concentrated near the origin,
which represents the colors of the photospheres of normal stars.
The object distribution extends to the upper-right direction
for stars with warm circumstellar dust.
In the diagrams including the [J]$-$[K] axis
(i.e. Figs.~\ref{fig:cc_density}a and \ref{fig:cc_density}c),
[K]--excess objects are mixed with
objects reddened by interstellar extinction.
In the diagrams including the [K]$-$[9] axis
(i.e. Figs.~\ref{fig:cc_density}a and \ref{fig:cc_density}b),
very red ([K]$-$[9]$>$5) objects,
presumably dominated by hot dust emission or
the emission of polycyclic aromatic hydrocarbons (PAHs),
can be clearly distinguished
from the majority of the objects,
although they are relatively small in number.

We define the occupation zone
of each class exclusively on these color-color diagrams
based on the locations of the known members.
The defined zones are overlaid in
Figs.~\ref{fig:ccds1}, \ref{fig:ccds2}, and \ref{fig:ccds3}.
%
%
Table~\ref{tbl:cc} summarizes
the numbers of known members used in the definition of the occupation zones,
and the total number of the sources
in the corresponding zones.
The occupation zones defined in this work are 
consistent with the criteria used in the source selections
in the related works using AKARI/MIR survey data
\citep[][]{Yita,Takita}.

\begin{table*}[h]
\caption{Classification of the AKARI/MIR catalogue sources.\label{tbl:cc}}
\begin{tabular}{clrrrrrr}\hline\hline
Group&Representative population &
\multicolumn{6}{c}{Number of members}\\
&(Definition in SIMBAD\tablefootmark{a})&
\multicolumn{2}{c}{[J]--[K] vs. [K]--[9]}&
\multicolumn{2}{c}{[K]--[9] vs. [9]--[18]}&
\multicolumn{2}{c}{[J]--[K] vs. [9]--[18]}\\
&&
known\tablefootmark{b}&
total\tablefootmark{c}&
known\tablefootmark{b}&
total\tablefootmark{c}&
known\tablefootmark{b}&
total\tablefootmark{c}\\\hline
A  &YSOs (YSO, TT$*$, HII)                   &  1,302& 81,802&   776&17,369 &   776&  5,588\\
B  &Emission line stars (Be$*$,Em$*$,WR$*$)  &  1,984&  1,644&   744&-\tablefootmark{d}&744&1,493\\
C  &Normal stars (O--M dwarfs \& giants; $*$)&167,829&288,597&35,117&36,906 &35,117& 27,836\\
D  &S-stars (S$*$)                           &    852&205,255&   461&17,454 &   461& 20,973\\
E  &C-rich AGBs (C$*$)                       &  5,197&202,252& 3,037&13,944 & 3,037& 18,596\\
F  &O-rich AGBs (OH$*$, AB$*$)               &  1,193& 30,711&   745&71,190 &   745& 83,012\\
G  &Post-AGBs \& PNe (pA$*$, PN)             &    836&  6,858&   670& 1,762 &   670&  2,076\\
H  &Galaxies (G, EmG, H2G, rG, LSB,          &  1,323&  2,735&   744& 3,977 &   744&  3,028\\
   &\hspace{1.4cm}LIN, Sy2, Sy1, AGN, SyG)   &       &       &      &       &      &       \\\hline
\end{tabular}
\tablefoot{
\tablefoottext{a}{Abbreviated expressions for object types
used in the SIMBAD database.
We assign ambiguous objects to each class.
For example, the objects labeled as C$*$? (carbon star candidate)
are included in group E.}
\tablefoottext{b}{The number of known objects in the SIMBAD database.}
\tablefoottext{c}{The number of objects in the occupation zone 
defined in this work.}
\tablefoottext{d}{Emission-line stars cannot be separated clearly on
the [K]--[9] vs. [9]--[18] diagram.}
}
\end{table*}

\begin{figure}[h]
\center
\includegraphics[width=7.5cm]{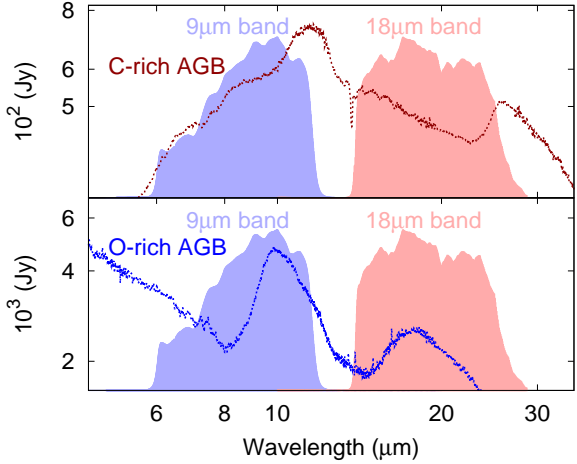}
\caption{Relative response curves of the
AKARI 9$\mu$m band and 18$\mu$m band.
The ISO/SWS spectra of 
typical Galactic C-rich and O-rich stars are overlaid
as references \citep{Sloan}.
An example of the C-rich star, IRC +40540,
displays carbonaceous 
features around 11 and 30\,$\mu$m,
that are not covered by the AKARI bands.
A typical O-rich star, $o$Cet, shows silicate features
around 10 and 18\,$\mu$m, 
which are efficiently covered by the AKARI bands.
\label{fig:rsr}}
\end{figure}

The C-rich and O-rich AGB stars can be clearly
distinguished in all the three 
color-color diagrams
(Figs.~\ref{fig:ccds1}a,~\ref{fig:ccds2}a, and ~\ref{fig:ccds3}a).
This is because 
the silicate features of O-rich AGB stars
around 10 and 20\,$\mu$m are
well-covered by the AKARI 9\,$\mu$m and 18\,$\mu$m bands,
while the carbonaceous (e.g. SiC) features of C-rich AGB stars
do not contribute to these AKARI bands (Fig.~\ref{fig:rsr}).
These characteristics are effective in helping us to differentiate
C-rich from O-rich AGB stars, 
and their availability is
one of the unique advantages 
of the AKARI/MIR survey data
compared with the other infrared surveys of IRAS \citep{IRAS},
MSX \citep{MSX}, and WISE \citep{WISE}.
The distinction of C-rich from O-rich objects is
clearest in the [J]$-$[K] vs. [9]$-$[18] color-color
diagram (Fig.~\ref{fig:ccds3}a).
Thus, we first use the classification 
in the [J]$-$[K] vs. [9]$-$[18] color-color diagram
(hereafter called candidate samples),
and then 
treat the sources residing
in the occupation zones
in all the three color-color diagrams
as more purified sub-samples
(hereafter called purified samples).
As a result,
the candidate samples contain 18,596 C-rich and 
83,012 O-rich AGB stars, while 
the purified samples contain 5,537
C-rich and 11,416 O-rich AGB stars (Table~\ref{tbl:pure}).

As described above,
contamination is inevitable for this type of method.
We investigate
the reliability and 
the completeness of the classified samples as follows
and summarize them in Table~\ref{tbl:pure}.
The samples located in each occupation zone
are composed of four types of objects:
(a) correctly classified known objects;
(b) known objects 
that are classified as other categories,
but can be regarded as the same group;
(c) unknown objects; and
(d) known objects that should belong to another group.
The objects labeled as variable stars and infrared objects are included
in (b) for C-rich and O-rich stars.
Mira-type stars are also included in (b) for O-rich stars.
For each occupation zone,
we define
($n_{\rm a}+n_{\rm b})/(n_{\rm a}+n_{\rm b}+n_{\rm c}+n_{\rm d})$ 
as a lower limit,
and ($n_{\rm a}+n_{\rm b}+n_{\rm c})/(n_{\rm a}+n_{\rm b}+n_{\rm c}+n_{\rm d})$
as an upper limit to the reliability
of the classified samples,
where $n_{\rm a}$, $n_{\rm b}$, $n_{\rm c}$, and $n_{\rm d}$ are
the numbers of objects in (a), (b), (c), and (d), respectively.
For each object class,
the completeness is estimated
as the ratio of
the number of correctly classified known objects
to the total number of known objects
of the same class in the AKARI MIR PSC.
The number of known objects for each class
depends on the previous incomplete surveys.
However, that is only the method to assess our samples,
that is useful to judge the validity of the statistical results
based on our samples.
For the candidate samples,
the reliability is 44--95\% for C-rich AGB stars,
and 53--95\% for O-rich AGB stars.
The completeness is 82\% for C-rich AGB stars
and 80\% for O-rich AGB stars.
For the purified samples,
the reliability is 68--96\% for C-rich AGB stars,
and 71--96\% for O-rich AGB stars.
The completeness is 64\% for C-rich AGB stars and 29\%
for O-rich AGB stars.
The completeness for O-rich AGB stars
is lower than for C-rich AGB stars.
As shown in Figures \ref{fig:ccds1}, \ref{fig:ccds2}, and \ref{fig:ccds3}, 
since the O-rich and C-rich samples are bluer,
they are more intermixed,
whereas the redder ones in the same samples can be clearly separated.
Therefore, with the present classification, it is inevitable that
the completeness for the redder members is higher than 
that for the bluer members.
Thus, the completeness of the C-rich samples
is higher than that of the O-rich samples, 
which contain a larger numbers of lower-mass non-dusty stars.

\begin{table*}[h]
\caption{Reliability 
and completeness of the newly classified C-rich and O-rich samples.
\label{tbl:pure}}
\begin{tabular}{crccrcc}\hline\hline
Selection &
\multicolumn{3}{c}{C-rich AGBs}&
\multicolumn{3}{c}{O-rich AGBs}\\
&Number& Reliability\tablefootmark{a} & Completeness\tablefootmark{b}
&Number& Reliability\tablefootmark{a} & Completeness\tablefootmark{b}\\
&& (\%) &  (\%) && (\%)&  (\%)\\\hline
$[J]-[K]$ vs. $[K]-[9]$ & 202,253 &19--96  &  78 & 30,712 & 48--92 &   30   \\
$[K]-[9]$ vs. $[9]-[18]$& 13,945 &55--89  &  85 &  71,191 & 52--94 &   71   \\
$[J]-[K]$ vs. $[9]-[18]$\tablefootmark{c}& 18,597 &44--95  &  82 &  83,013 & 53--95 &   80   \\
All the diagrams\tablefootmark{d}&  5,537 &68--96  & 64 &  11,416 & 71--96 & 29 \\\hline
\end{tabular} 
\tablefoot{
\tablefoottext{a}{The upper limit
is calculated for the extreme case that
the classification of all the unknown objects located in each occupation zone
is true,
and the lower limit is calculated
for the extreme case that the classification of all the unknown objects 
in each occupation zone is false.}
\tablefoottext{b}{The ratio of
the number of correctly classified known objects
to the total number of the known objects in
each of C-rich and O-rich AGBs.}
\tablefoottext{c}{Candidate samples.}
\tablefoottext{d}{Purified samples
residing in the occupation zones
in all the three color-color diagrams (see Sect.~2).
}
}
\end{table*}

\section{Results} \label{result}

\subsection{Galactic distributions}

\begin{figure*}[h]
(Group A) YSOs\hspace{3.8cm}
(Group C) Normal stars\hspace{2.7cm}
(Group D) S-stars\\
\includegraphics[width=6.2cm]{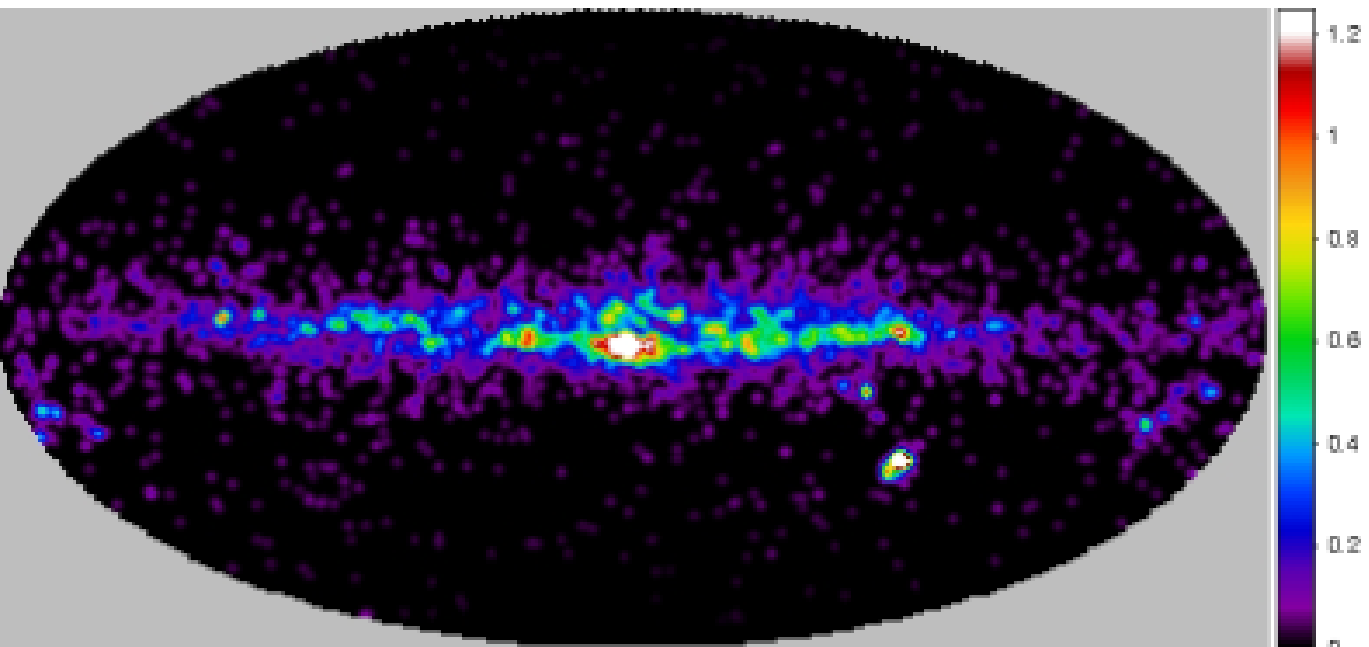}
\includegraphics[width=6.2cm]{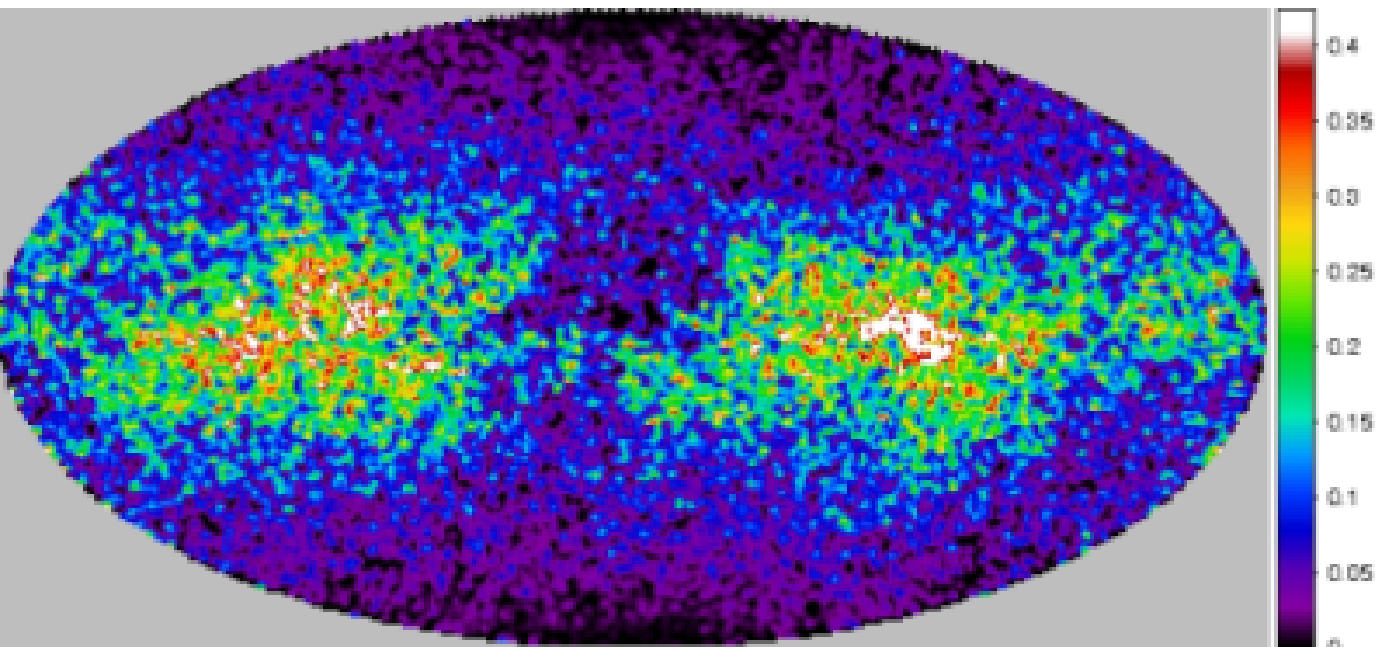}
\includegraphics[width=6.2cm]{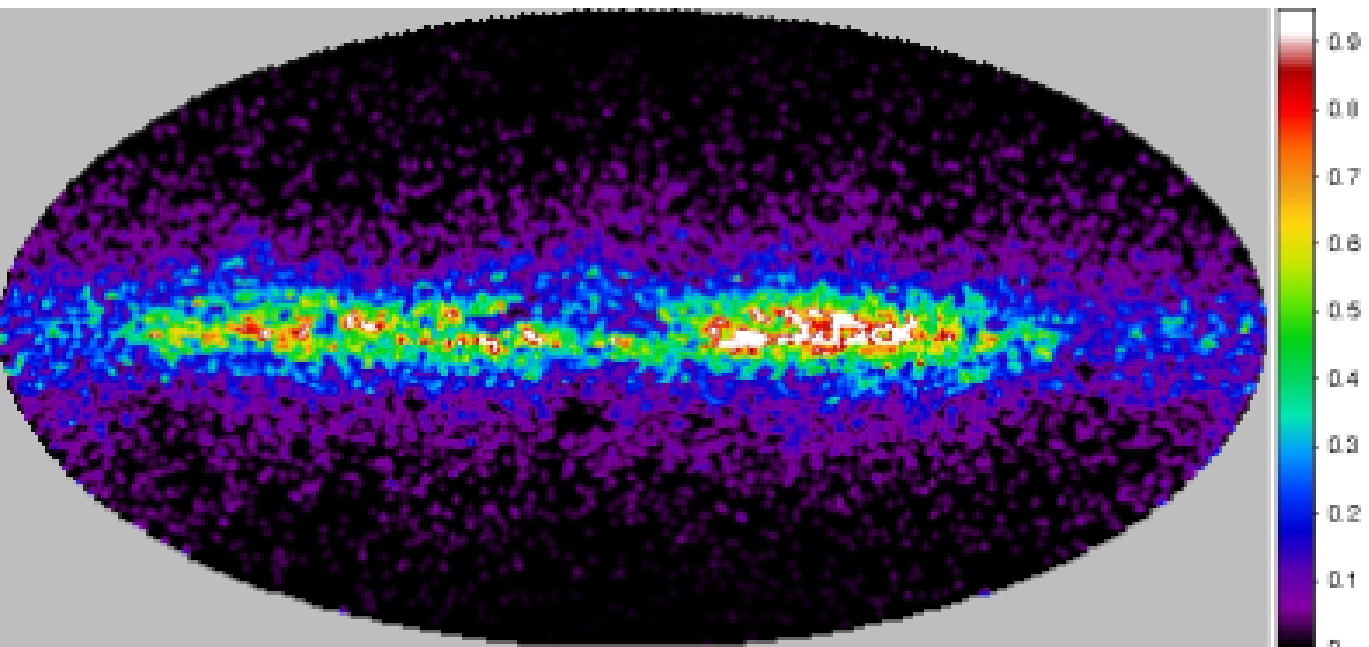}\\
(Group E) C-rich AGBs\hspace{2.8cm}
(Group F) O-rich AGBs\hspace{2.8cm}
(Coordinates)\\
\includegraphics[width=6.2cm]{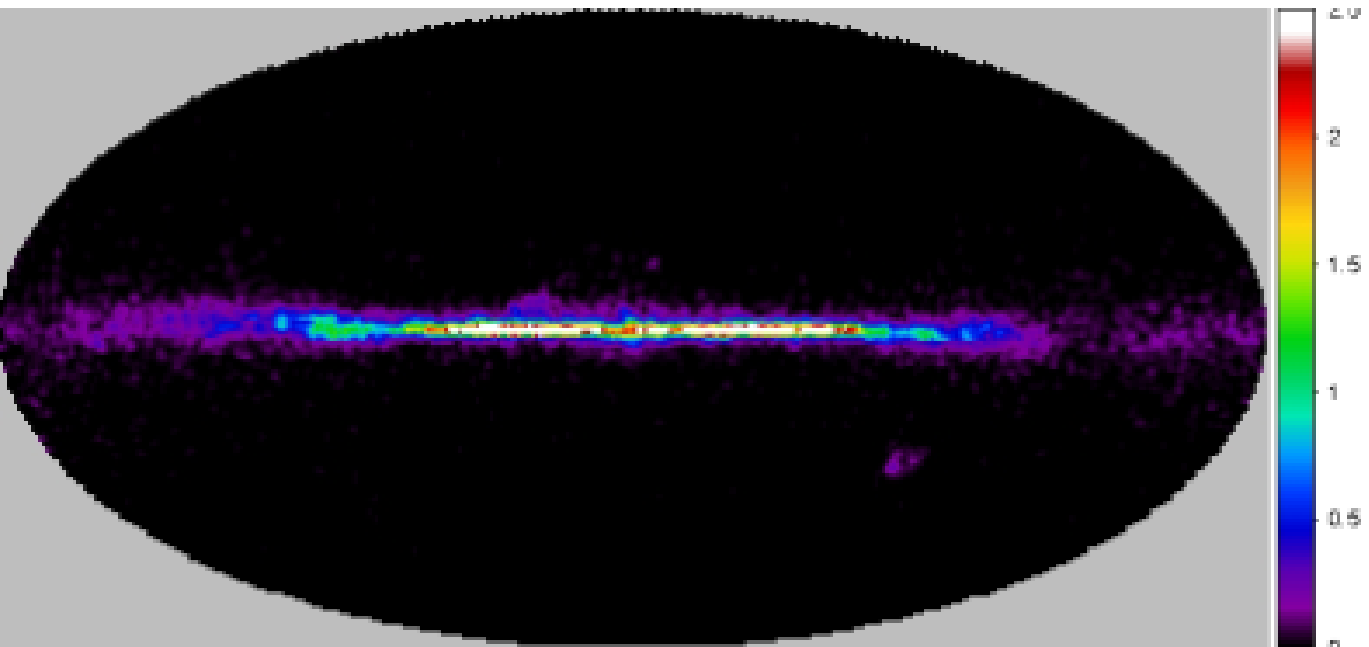}
\includegraphics[width=6.2cm]{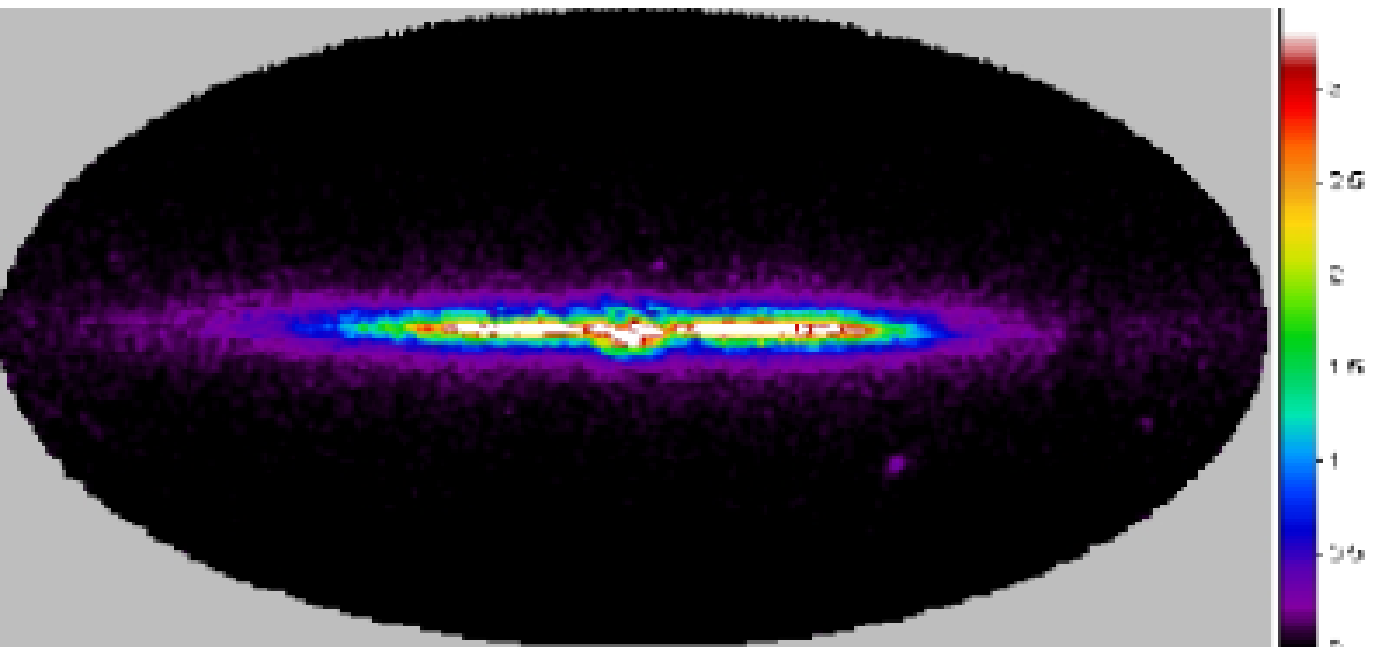}
\includegraphics[width=6.2cm]{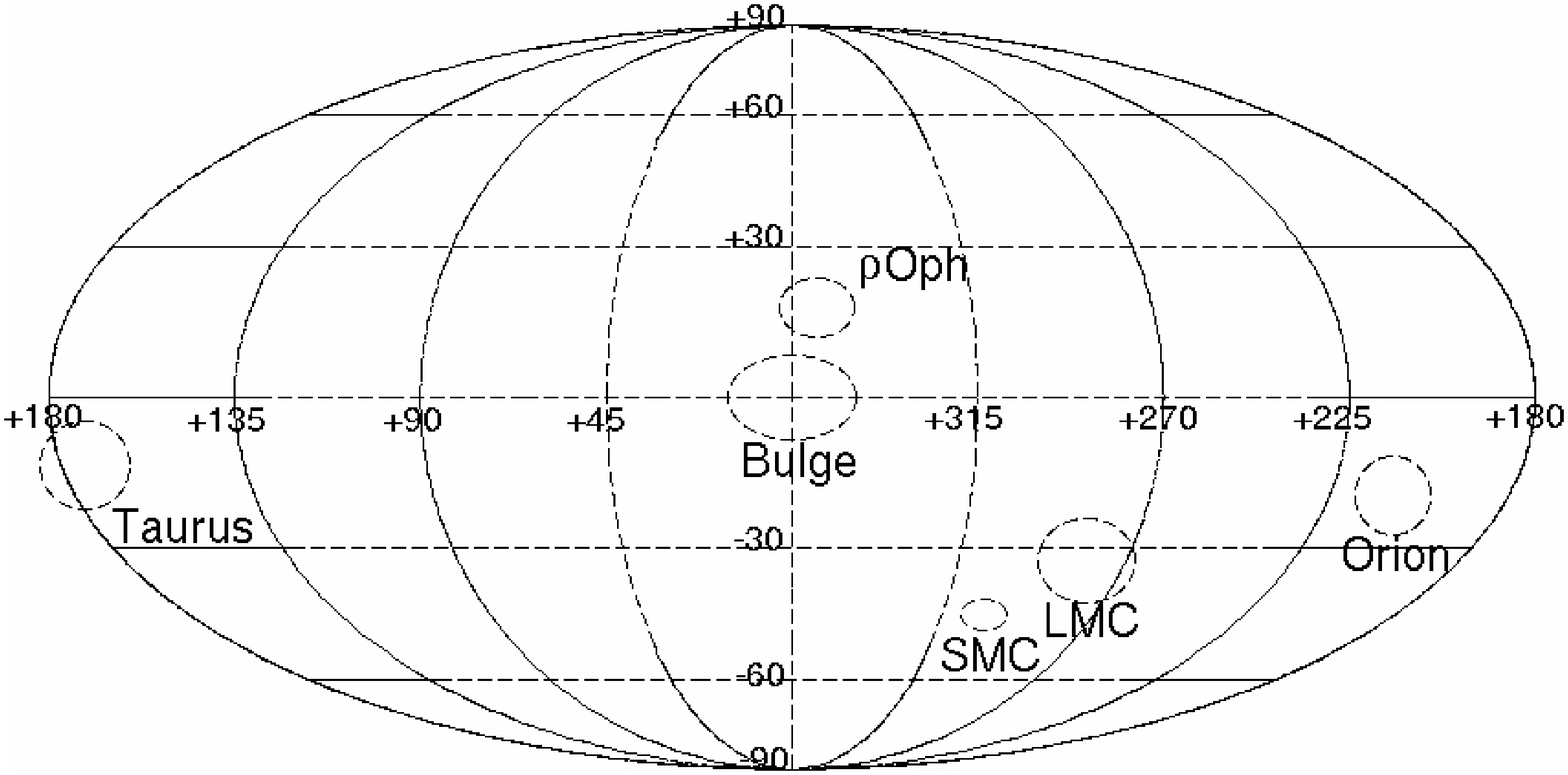}\\
\caption{Spatial distributions of the AKARI/MIR sources
in density maps
in the Galactic coordinates of the Aitoff projection.
These sources are classified
on the basis of the 
[9]$-$[18] vs. [J]$-$[K] color-color diagram alone
(candidate samples).
We show groups A, C, D, E, and F
because they have a sufficiently large number of members.
The color scales 
are linear from 0\% to 50\% of the peak value
and given in units of deg$^{-2}$.
The coordinate grids and the locations of
the Galactic bulge, the LMC, the SMC, and 
the nearby star-forming regions
are shown in the {\em lower right} panel.
\label{fig:galactic}}
\end{figure*}

Figure~\ref{fig:galactic} shows
the spatial distributions of the objects selected
in the [9]$-$[18] vs. [J]$-$[K] color-color diagrams
in density maps in the Galactic coordinates.
The surface densities in the maps in
Fig.~\ref{fig:galactic}
are calculated for the area of 1$\times$1 deg$^2$,
and given in units of deg$^{-2}$.
The spatial distributions of S-stars,
C-rich AGB stars, and
O-rich AGB stars
trace the Galactic disk.
%
%
The bulge, the LMC, and the Small Magellanic Cloud (SMC)
are also recognized in these maps.
Dwarfs and giants
(group C in Fig.~\ref{fig:galactic})
show relatively uniform distributions
in contrast to the other populations
except for two dim concentrations
around $l = -90^\circ$ and $+90^\circ$.
Since normal stars are faint in the MIR bands,
their detection volume is limited to
the local space ($< 1$\,kpc) around Sun.
The distributions are similar to that of 
the Hipparcos stars \citep{Hip},
corresponding to the structure of the local arm.
In the panel for YSOs 
(group A in Fig.~\ref{fig:galactic}),
nearby star-forming regions such as Orion, Taurus, and $\rho$Oph
are recognized.
These maps 
demonstrate the plausibility of the classification
by reproducing well-known major characteristics
of the spatial distribution of each Galactic object.

\subsection{Distributions of C-rich and O-rich AGB stars}

\begin{figure*}[h]
(a) C-rich AGBs (purified samples)\hspace{4cm}
(b) O-rich AGBs (purified samples)\\
\includegraphics[width=9.3cm]{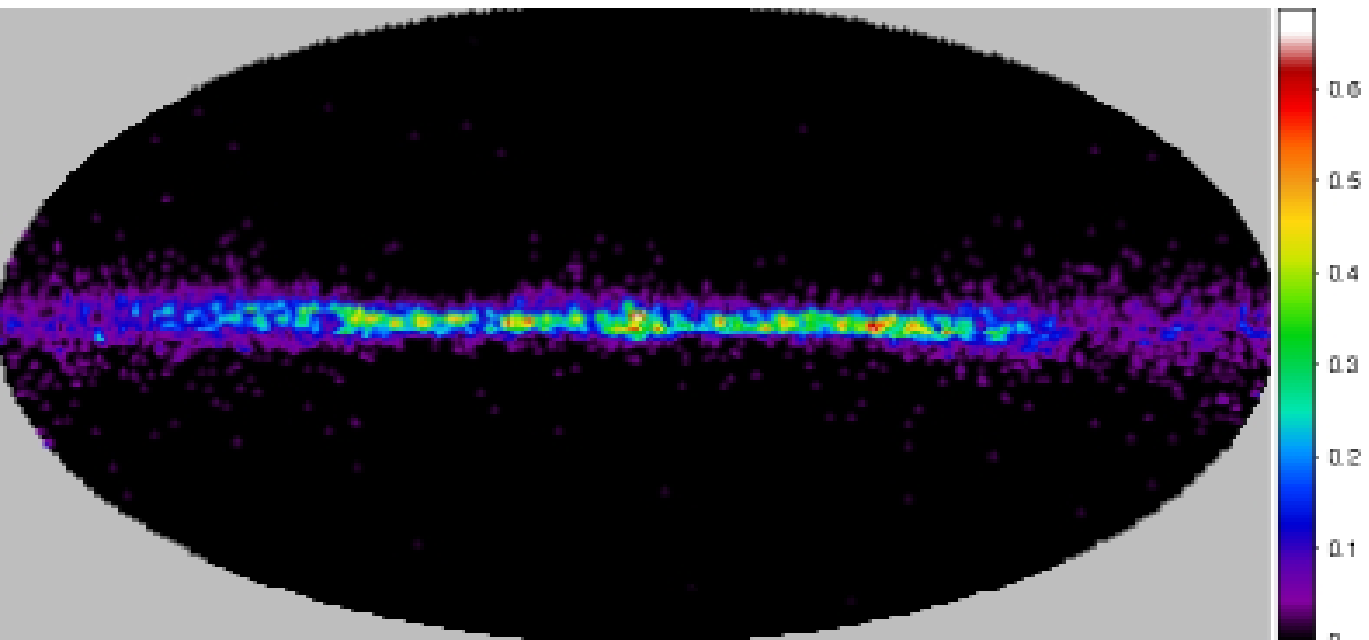}
\includegraphics[width=9.3cm]{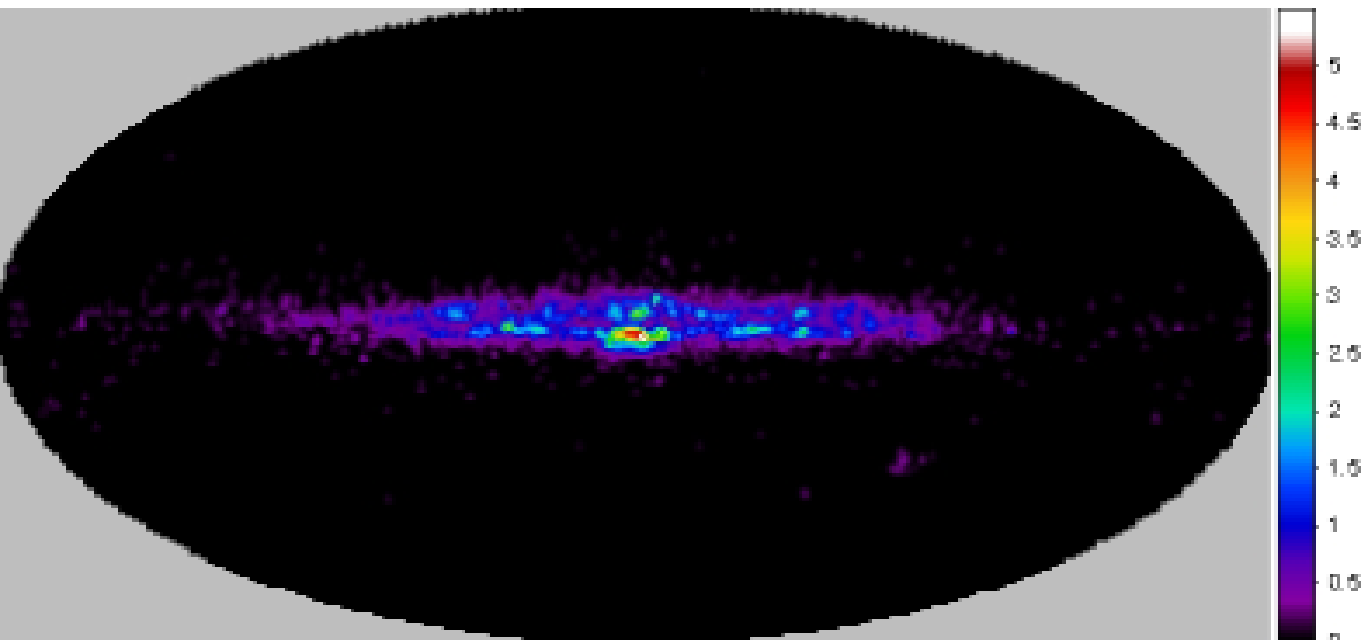}
\caption{Spatial distributions of the purified samples of 
(a) C-rich AGB stars and (b) O-rich AGB stars
in density maps in the Galactic coordinates.
The purified sample corresponds to those belonging to 
all the occupation zones of
the [J]$-$[K] vs. [K]$-$[9] (Fig.~\ref{fig:ccds1}a),
[K]$-$[9] vs. [9]$-$[18] (Fig.~\ref{fig:ccds2}a), and
[J]$-$[K] vs. [9]$-$[18] diagrams (Fig.~\ref{fig:ccds3}a).
The color scales 
are linear from 0\% to 100\% of the peak value
and given in units of deg$^{-2}$.
\label{fig:CandO}}
\end{figure*}

\begin{figure*}[h]
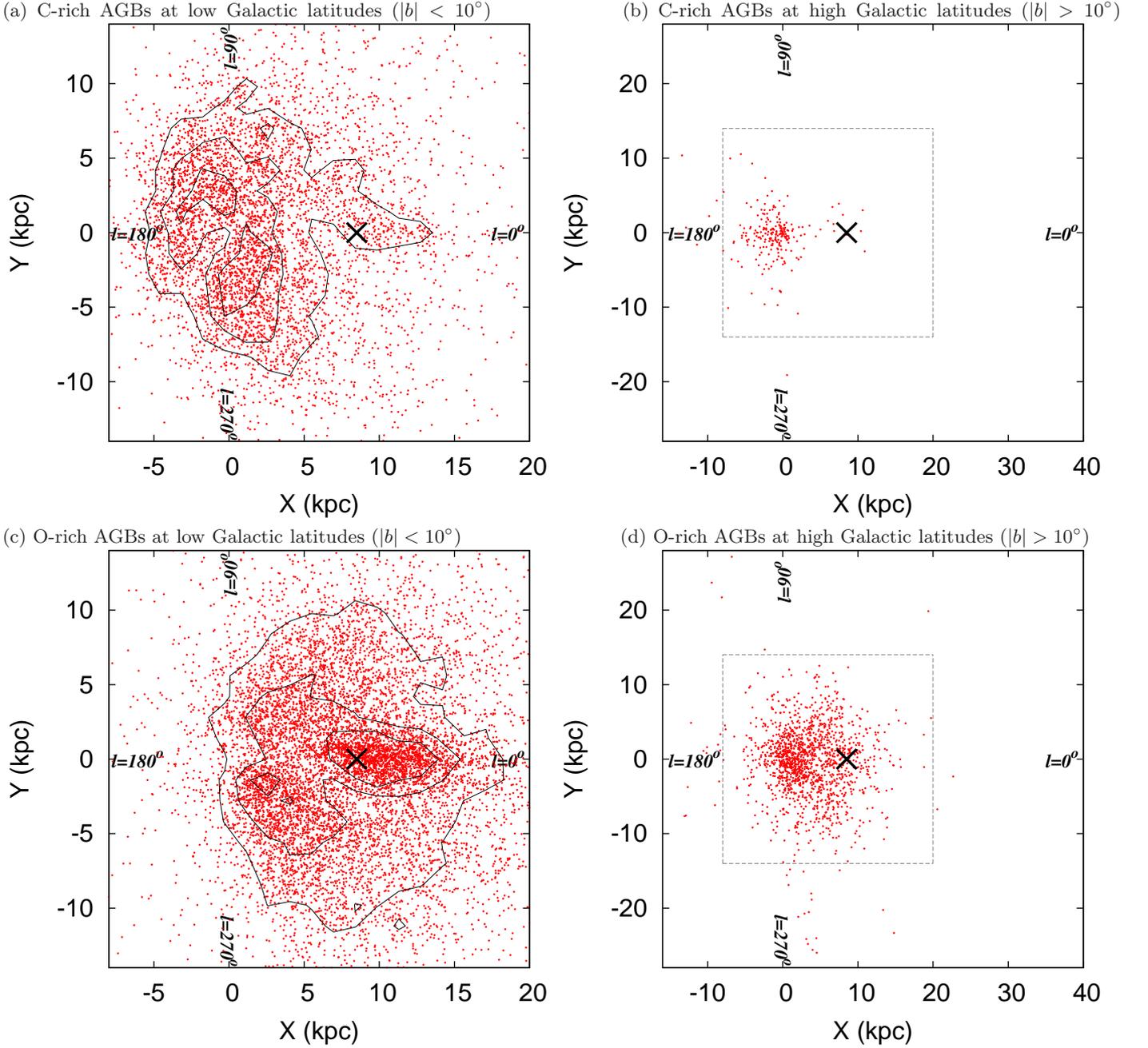

(a) C-rich AGBs at low Galactic latitudes ($|b|<10^\circ$)\hspace{2cm}
(b) C-rich AGBs at high Galactic latitudes ($|b|>10^\circ$)
\includegraphics[width=9cm]{circC2.epsi}
\includegraphics[width=9cm]{circCout2.epsi}\\
(c) O-rich AGBs at low Galactic latitudes ($|b|<10^\circ$)\hspace{2.5cm}
(d) O-rich AGBs at high Galactic latitudes ($|b|>10^\circ$)\\
\includegraphics[width=9cm]{circO2.epsi}
\includegraphics[width=9cm]{circOout2.epsi}
\caption{Spatial distributions
of the C-rich and O-rich AGB stars
projected onto the Galactic plane viewed from the north pole:
(a) C-rich AGBs at $|b|\leq10^\circ$,
(b) C-rich AGBs at $|b|>10^\circ$,
(c) O-rich AGBs at $|b|\leq10^\circ$,
and (d) O-rich AGBs at $|b|>10^\circ$.
%
The red dots indicate the purified samples.
The contours in panels (a) and (c) show
the areal densities for the purified samples
(i.e. densities of red points).
The levels of the contours are
10, 20, and 30 kpc$^{-2}$ for panel (a),
and 10, 50, and 100 kpc$^{-2}$ for panel (c).
The distance scale for each object is estimated from
the AKARI 9\,$\mu$m fluxes and [K]$-$[9] colors (see Sect.~3).
The Sun is located at the origin.
The G.C. is to the right of the Sun
and shown by the cross at $(X,Y)=(8.5,0)$.
%
%
The dotted squares in (b) and (d) indicate the size of (a) and (c).
Note that the scale of the distance used in panels (b) and (d)
is a factor of two larger than that in panels (a) and (c).}
\label{fig:proj}
\end{figure*}

Figures~\ref{fig:CandO}a and \ref{fig:CandO}b show
the celestial distributions
of the purified C-rich and O-rich AGB samples, respectively,
as density maps in Galactic coordinates.
By assuming that
the star counts can be described using Poisson statistics,
the 2-$\sigma$ significance levels of 
{the surface densities of the C-rich and O-rich AGB samples are
0.3 and 1.5 deg$^{-2}$, respectively.
The O-rich stars show
a concentration toward the G.C. 
(the areal density varies from 3\,deg$^{-2}$ in the G.C.
to $<$0.5\,deg$^{-2}$ in the outer Galaxy ($|l|>90^\circ$)),
whereas C-rich stars show a uniform distribution
along the Galactic plane 
(the areal density is kept to within
0.2--0.4 deg$^{-2}$ across the Galactic plane).
Thus, the decrease in density along the Galactic plane
seen in Fig.~\ref{fig:CandO}b is significant, 
while the small-scale structures in Fig.~\ref{fig:CandO}a are not. 
These results can be accounted for either
by an intrinsic difference in the distribution
on a Galactic scale, or 
by a situation where O-rich stars are distributed
on a Galactic scale in the thin disk
with interstellar extinction and 
C-rich stars are located in relatively nearby regions.
To investigate any difference in greated detail,
three-dimensional information would be indispensable.

We plot
C-rich and O-rich AGB stars 
in the projection onto the Galactic plane
of both disk stars ($|b|<10^\circ$; Figs.~\ref{fig:proj}a and c)
and stars at high Galactic latitudes
($|b|>10^\circ$; Figs.~\ref{fig:proj}b and d).
The distances of the individual samples 
were estimated
from their AKARI 9\,$\mu$m fluxes ($F_{\rm 9\mu m}$)
and [K]$-$[9] colors.
A formula to estimate the distances 
was empirically derived using the sub-samples
for which 
the distance and the mass-loss rates were known.
These parameters were estimated
based on MIR observations independently
by \citet{Zhang} and \citet{IRTS2}.
We derived the relation
among the distance, the mass-loss rates, and $F_{\rm 9\mu m}$
for the sub-samples.
We also derived the relation 
between the mass-loss rates and [K]$-$[9] colors
(see Appendix for details of the derivation).
By combining both relations, 
we obtained a formula
to derive the distance from $F_{\rm 9\mu m}$ and [K]$-$[9].
The maximum distance that we can probe 
in our sample is limited 
by the detection limit for the AKARI 18\,$\mu$m flux.
The maximum distance to
the faintest members in our sample
is estimated to be 8$\pm$2\,kpc,
based on the above flux-distance relation 
with typical [9]$-$[18] colors (also see Appendix). 

The edge of the outer part of our Galaxy
(the left side of Fig.~\ref{fig:proj}) indicates
the end of the corresponding stellar populations,
whereas 
the edge toward
the direction of the opposite side over the G.C.} 
%
%
(the right side of Fig.~\ref{fig:proj})
corresponds to
the detection limit of this survey for the corresponding objects.
%
The elliptical concentration is seen around the G.C.
in the panels for the disk component ($|b|<10^\circ$;
Figs.~\ref{fig:proj}a and \ref{fig:proj}c).
The inclination of about five degrees to the Sun--G.C. axis
may indicate the central bar \citep{Gerhard,Cole}.
However, we note that
the elongated distribution is also
caused by the uncertainties 
in the derived distances to the individual stars.
%
%
The concentrations of sources around $l\sim280$--$300^\circ$
in the panels for high latitude objects
($|b|>10^\circ$; Figs.~\ref{fig:proj}b and \ref{fig:proj}d)
indicate the LMC and the SMC.
They are clearly separated from our Galaxy,
though there remains 
an elongated distribution towards the LMC
presumably caused by the errors in the distance estimates.

The most remarkable finding is that
O-rich AGB stars are
concentrated toward the G.C.
while C-rich AGB stars have a relatively uniform distribution 
across the Galactic disk.

\section{Discussion} \label{discussion}

\begin{figure}[h]
\includegraphics[width=9cm]{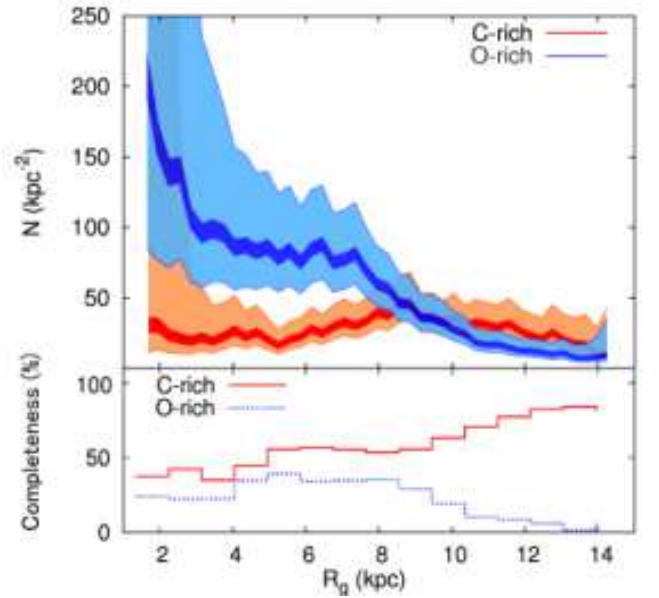}
\caption{{\em Upper panel}: areal densities
of the number of the purified C-rich (red)
and the O-rich (blue) AGB samples
in the Galactic plane ($|b|<10^\circ$)
as a function of Galactocentric distance ($R_g$),
not corrected for the completeness (see below).
For each $R_g$,
we count the numbers of the disk AGB stars
included in the annular regions of width 0.3\,kpc
within 8\,kpc of Sun
(the penetrating depth of our sample)
and divide the star counts by the corresponding areas. 
The narrower stripes indicate
the errors estimated from the Poisson statistics for the star counts, 
while the wider stripes indicate
the ones derived by assuming uncertainties
of $\pm$2\,kpc in estimating the distance. 
%
{\em Lower panel}: completeness of both samples
as a function of the Galactocentric distance ($R_g$).
\label{fig:hist}}
\end{figure}

To have
more quantitative discussion about the difference in the
spatial distributions of C-rich and O-rich AGB stars,
we plot the areal number density of our samples
as a function of Galactocentric distance ($R_g$) in Fig.~\ref{fig:hist}.
At various galactocentric distances,
we count the numbers of disk AGB stars
included in the annular regions of width 0.3\,kpc
within 8\,kpc of the Sun considering the survey depth.
The smaller errors 
indicated by darkly colored stripes
in Fig.~\ref{fig:hist}
are estimated by the Poisson statistics in the star counts, 
while the larger ones 
indicated by more lightly colored stripes include an uncertainty
of $\pm$2\,kpc in estimating the distance. 
We also investigate the completeness of both samples
as a function of $R_g$ in the lower panel of Fig.~\ref{fig:hist}.
The completeness of the C-rich sample increases with $R_g$,
while that of the O-rich sample is lower in
the outer galaxy ($R_g>9$\,kpc).
For O-rich AGB stars, 
the completeness is higher for the redder members (see Sect.~2).
The lower completeness in outer Galaxy means that
the fraction of dusty members 
(i.e. massive O-rich AGB stars)
becomes smaller in the outer Galaxy.
In contrast, the completeness of the C-rich AGB stars shows
little depends on their color.
The number and variety of the catalogued C-rich AGB stars
are relatively small in the outer Galaxy. Thus, the completeness
of C-rich AGB stars 
increases toward the outer Galaxy.
The ratio of the completeness of the C-rich sample
to that of the O-rich sample 
is found to be almost constant at $R_g<9$\,kpc.
Hence, our result 
is not affected much by the variations in the completeness. 

Table~\ref{tbl:prev} summarizes 
the related investigations with
the numbers and the penetrating depths of the
samples. 
Previous studies have consistently reported
relatively flat distributions of C-rich AGB samples
compared with O-rich AGB samples,
which are instead concentrated toward the G.C.
Our result in Fig.~\ref{fig:hist} 
quantitatively confirms the trends reported in
the previous works, which improves
the spatial completeness
by extending the survey volume to 8\,kpc from the Sun
and avoiding source confusions toward the G.C.

\begin{table*}
\caption{Related investigations
of the spatial distribution of C-rich AGBs.\label{tbl:prev}}
\begin{tabular}{lrrl}\hline\hline
References      & Number  &Distance& Sample selection\\
                &of sample&(kpc)   & \\\hline
\citet{Claussen}& 215     & 1.5    & TMSS photometries and optical spectra\\
\citet{Thronson}& 619     & 5      & IRAS colors\\
\citet{Jura}    & 126     & 2.5    & Optical, TMSS, and IRAS colors\\
\citet{Noguchi} & 2,250\tablefootmark{a}   &3.6  & IRAS colors\\
This work (2010)& 18,597\tablefootmark{b}&8    & AKARI/MIR colors\\
                & 5,537\tablefootmark{c}&8    & AKARI/MIR colors\\\hline
\end{tabular}
\tablefoot{
\tablefoottext{a}{Contamination is estimated to be 40\%.}
\tablefoottext{b}{Candidate sample.}
\tablefoottext{c}{Purified sample.}
}
\end{table*}

Why do the distributions
of C-rich and O-rich AGB stars differ ?
One possible reason is that
the metallicity of the ISM at their birth place
affects the stellar chemical composition of the photospheres
\citep[e.g.][]{Noguchi}.
%
The metallicity gradient along $R_g$
was reported by many authors
(e.g. by using Cepheids and open clusters).
The reported gradient of $d{\rm [Fe/H]}$/$dR_g$ at $R_g$ = 5--15\,kpc
varies from $-$0.04 to $-$0.13 dex kpc$^{-1}$ \citep{Pedicelli},
and is
higher toward the inner Galaxy.
The [O/H] gradients are also reported by several authors
\citep[e.g.][]{Smartt, Shaver, Rudolph}, to be between about
$-0.04$ and $-0.07$ dex kpc$^{-1}$ at radii 6--18\,kpc.
The log(C/M) value at radii $R_{g}=$2 to 14\,kpc
in our result has
the gradient of 
$\sim$0.1\,dex kpc$^{-1}$,
which has an inverse gradient of slope
similar in absolute value to that of the metallicity.

The anti-correlation between 
the metallicity and the C/M ratio is also reported for nearby galaxies.
\citet{Brewer} reported on M31, a spiral galaxy similar to our Galaxy,
observing five regions located at 4--32\,kpc from the G.C.
The metallicity decreases along the $R_g$,
whereas the C/M ratio increases.
\citet{Pritchet} confirmed that
the number of C-rich AGB stars per unit luminosity shows
a correlation with the average [Fe/H] among the galaxies,
IC1613, SMC, NGC6822, NGC55, M33, LMC, NGC300, M31, and our Galaxy,
thus the lower metallicity results in the higher C/M ratio.
The relation is approximated by 
$d({\rm [Fe/H]})/d(\log{\rm (C/M)})\sim-0.5$ \citep[e.g.][]{Battinelli,Cioni}.
Our result in Fig.~\ref{fig:hist} shows that
the slope is $d({\rm [Fe/H]})/d(\log{\rm (C/M)})= -0.4\sim-1.3$
at R$=$2--14\,kpc,
which connects smoothly with the value for nearby galaxies.
Thus, the trend reported in nearby galaxies
is applicable to our Galaxy, 
while the ranges of metallicity are different:
$-$1.3$<$[Fe/H]$<$0 for nearby galaxies,
and $-$0.5$<$[Fe/H]$<$1 for our Galaxy.
We note that
the metallicity dependence of the C/M ratio
in nearby galaxies is based on 
optical \citep{Battinelli} or
near-infrared \citep[NIR;][]{Cioni} observations,
whereas our result is based on MIR observations.
Figure \ref{fig:hist} also suggests that C/M$\sim 1$ at [Fe/H]$=$0
(i.e. the solar abundance), 
although our O-rich samples are biased toward dusty members. 

How does the difference in spatial distributions
between C-rich and O-rich AGB stars affect
the ISM in our Galaxy ?
Our result indicates that as dust suppliers
the population of C-rich stars
are minor contributers in the inner Galaxy
but major contributors around the Sun
and in the outer Galaxy.
It is reasonable to assume that 
the mass loss from AGB stars 
has a direct effect on
the chemical composition of their local ISM.
The spatial constraint on the dominant replenishment
of carbonaceous grains to the ISM
may have implications
for habitable zones in our Galaxy \citep{GHZ},
though a significant theoretical gap remains
between the supply of carbonaceous grains
to the interstellar space
and the origin of organic material.
The spatial variation in
the amount of interstellar carbonaceous grains
relative to that of silicate grains
has not been reported on the Galactic scale.
Furthermore, it is well-known that
the PAH (as a representative of carbonaceous grain) 
emission displays a
good spatial correlation
with the cold dust emission \citep[e.g.][]{OnakaIRTS,PAHdust,Bend};
the latter is mostly emitted by silicate grains.
We plan to perform
detailed and quantitative comparisons of 
the spatial distributions of 
carbonaceous and silicate grains
with
those of C-rich and O-rich AGB stars
in our Galaxy in future work.

\section{Summary}

On the bassis of AKARI observations,
we have investigated 
the spatial distributions
of C-rich and O-rich AGB stars in our Galaxy.
The C-rich and O-rich AGB stars
in the AKARI 9\,$\mu$m and 18\,$\mu$m All-Sky Survey catalogue
are well-distinguished 
by the empirical approach using color-color diagrams.
The reliability and the completeness
of the sample selection are discussed.
The C-rich AGB stars are
uniformly distributed within 8\,kpc from Sun,
while O-rich AGB stars are concentrated toward the G.C.
Our result confirms those of previous investigations,
extends the survey volume to the Galactic scale ($\sim$8\,kpc),
and enhances the completeness toward the inner Galaxy
by spatially resolving individual sources.
We also propose that
the metallicity gradient with galactocentric distance can be
one of the possible origins of the difference 
in the spatial distributions
of C-rich and O-rich AGB stars.

\begin{acknowledgements}
This work was supported by
the the Nagoya University Global COE Program, 
``Quest for Fundamental Principles in the Universe (QFPU)''
from JSPS and MEXT of Japan.
This research is based on observations with AKARI,
a JAXA project with the participation of ESA.
We thank all the members of the AKARI project.
This research has made use of the SIMBAD database,
operated at CDS, Strasbourg, France.
This publication makes use of data products 
from the Two Micron All Sky Survey,
which is a joint project of the University 
of Massachusetts and the Infrared Processing
and Analysis Center/California Institute of Technology,
funded by the National Aeronautics and
Space Administration and the National Science Foundation.
This research has made use of the NASA/IPAC
Infrared Science Archive,
which is operated by the Jet Propulsion Laboratory,
California Institute of Technology,
under contract with the National Aeronautics and Space Administration.
We also thank an anonymous referee for careful reading
and constructive comments.
\end{acknowledgements}


\clearpage
\appendix

\section{Distance estimate}
We first derive an empirical relation between
the distance ($D$;\,kpc),
the observed 9\,$\mu$m flux ($F_{\rm 9 \mu m}$;\,Jy),
and the [K]$-$[9] color
for well-studied AGB samples
\citep[][]{IRTS2,Zhang}.
We then apply this relation to our samples
and estimate $D$ from the observed values of
$F_{\rm 9 \mu m}$ and [K]$-$[9].
We use AKARI 9\,$\mu$m fluxes because they
are more acurate
than the 18\,$\mu$m fluxes.

The \citet{IRTS2} sample contains
126 C-rich stars and 563 O-rich stars
observed by the IRTS/NIRS.
The \citet{Zhang} sample contains
184 C-rich stars and 110 O-rich stars
selected from the literature.
Most members of both samples
have AKARI 9\,$\mu$m fluxes.
The mass-loss rates 
for both samples
are estimated observationally by using
methods independent of MIR observations,
while the distances are
estimated based on NIR and MIR fluxes.
The mixture of these samples (hereafter BZ sample)
comprehensively
covers both dust-poor members
($10^{-7.5}-10^{-5.5}$\,M$_\odot$yr$^{-1}$)
and dust-rich members
($10^{-5.5}-10^{-4}$\,M$_\odot$yr$^{-1}$).

Figure~\ref{fig:distance2} shows
the distances vs. the AKARI 9\,$\mu$m fluxes
for the BZ sample.
The observed flux is
inversely proportional to the square of the distance,
\begin{equation}
F_{\rm 9 \mu m} \propto L_{\rm 9\mu m}\times D^{-2} ,
\end{equation}
where
$L_{\rm 9 \mu m}$ is the 
intrinsic luminosity of a star at 9\,$\mu$m.
We assume that $L_{\rm 9 \mu m}$ is
a function of the amount of circumstellar dust,
thus a function of the mass-loss rate
($dM/dt$; ${\rm M}_\odot {\rm yr}^{-1}$) as
\begin{equation}
L_{\rm 9 \mu m} \propto (dM/dt)^\gamma,
\end{equation}
where $\gamma$ is a free parameter.

By combining Eqs. (A.1) and (A.2),
the distance is described as
a function of $F_{\rm 9 \mu m}$ and $dM/dt$ as
\begin{equation}
\log{D} = C_1 + C_2\log(dM/dt) - \frac{1}{2}\log(F_{\rm 9\mu m}) ,
\end{equation}
where $C_1$ and $C_2$ are
obtained as 6.1$\pm$0.06, and 0.35$\pm$0.01, respectively, 
by fitting the linear relation among 
$\log{(D)}$, $\log{(dM/dt)}$, and $\log{(F_{\rm 9\mu m})}$
for the BZ sample.

To obtain $dM/dt$ from our observations,
we assume that it is
a function of excess emission
indicated by the [K]$-$[9] color.
Figure~\ref{fig:massloss} shows 
the mass-loss rate
plotted against [K]$-$[9] color
for the BZ sample.
%
%
We assume that the dotted curve in Fig.~\ref{fig:massloss},
is an empirical relation between the mass loss rate and the color
expressed as
\begin{equation}
  \log(dM/dt) = \log(3.8\times(([K]-[9])-0.4))-8.0 .
\end{equation}

\begin{figure*}[H]
\hspace{5mm}
(a) C-rich AGB stars
\hspace{6cm}
(b) O-rich AGB stars\\
\includegraphics[width=8cm]{kekkaC3.epsi}
\hspace{5mm}
\includegraphics[width=8cm]{kekkaO3.epsi}
\caption{(a)
Heliocentric distance of C-rich AGB stars
as a function of the AKARI 9\,$\mu$m flux.
The open squares and the open circles show 
previously measured distances for well-studied samples
from \citet[][squares]{Zhang} and \citet[][circles]{IRTS2}, respectively.
The black dots show
distances estimated
for our purified samples.
The dashed curves indicate
the luminosity distance for the objects 
of [K]$-$[9]$=$14\,mag and [K]$-$[9]$=$1\,mag,
which correspond to 
the dust-rich case ($dM/dt\sim10^{-4}$ M$_\odot$yr$^{-1}$) and 
the dust-poor case ($dM/dt\sim10^{-8}$ M$_\odot$yr$^{-1}$),
respectively (see Fig.~\ref{fig:massloss}).
The vertical
lines indicate
the faint flux end of our sample (270$^{+150}_{-100}$\,mJy),
which corresponds to
the distance of 8$\pm2$\,kpc for the dust-poor members.
(b) The same plot as panel (a), but for O-rich AGB stars.
\label{fig:distance2}}
\end{figure*}

\begin{figure}[h]
\center
\includegraphics[width=8cm]{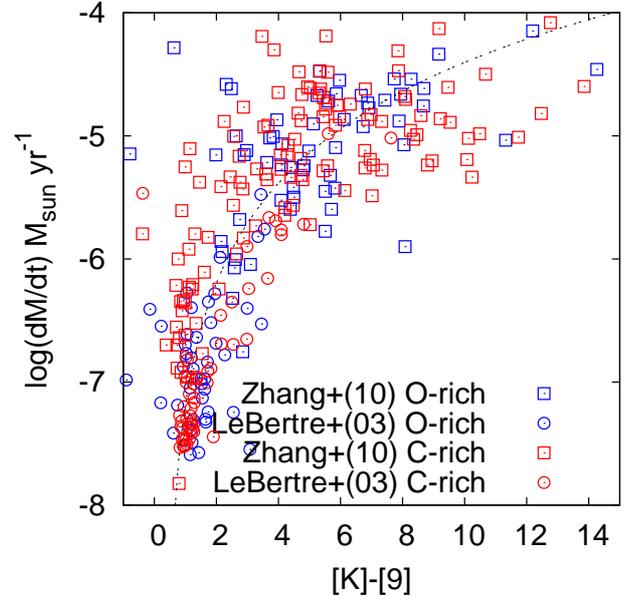}
\caption{Mass-loss rate of AGB stars
as a function of the [K]$-$[9] color,
measured by \citet{Zhang} and \citet{IRTS2}.
The symbols are the same as in Fig.~\ref{fig:distance2}.
\label{fig:massloss}}
\end{figure}

From equations (A.3) and (A.4) with
observed $F_{\rm 9\mu m}$ and [K]$-$[9],
we estimate the distances of our sample.
The results are overlaid with dots in Fig.~\ref{fig:distance2}.
The faintest $F_{\rm 9\mu m}$ of our sample is
limited by the detection limit of the AKARI 18\,$\mu$m band.
Typical [9]$-$[18] colors 
for dust-poor AGBs are 0.5$\pm0.5$\,mag
from Fig.~\ref{fig:ccds2}a.
By applying them to the flux detection limit at 18\,$\mu$m
\citep[90\,mJy;][]{MirCat},
the minimum $F_{\rm 9\mu m}$ for our sample is 270$^{+150}_{-100}$\,mJy,
which corresponds to the distance of 8$\pm2$\,kpc
from Fig.~\ref{fig:distance2}.
Figure~\ref{fig:distance2} also indicates that
the AKARI/MIR survey can probe
dust-rich AGB stars beyond 40\,kpc,
covering the objects in the LMC and the SMC.

\begin{thebibliography}{Mittelbach100}
\bibitem[Battinelli \& Demers (2005)]{Battinelli}	
Battinelli, P., \& Demers, S. 2005, \aap, 434, 657
\bibitem[Bendo et al.(2008)]{Bend}
Bendo, G. J., Draine, B. T., Engelbracht, C. W., et al.
2008, \mnras, 389, 629
\bibitem[Bertoldi et al.(2003)]{Bertoldi}
Bertoldi, F., Carilli, C. L., Cox, P., et al. 2003, \aap, 406, L55
\bibitem[Blanco(1965)]{Blanco}
Blanco, V. M. 1965 in Stars and Stellar Systems Vol. V,
ed. A. Blaauw, \& M. Schmidt (Chicago: University of Chicago Press), 241 
\bibitem[Brewer et al.(1995)]{Brewer}
Brewer, J. P., Richer, H. B., \& Crabtree, D. R. 1995, \aj, 109, 2480
\bibitem[Buchanan et al.(2009)]{Spitzercc} 
Buchanan, C. L., Kastner, J. H., Hrivnak, B. J., \& Sahai, R. 2009, \aj, 138, 1597
\bibitem[Chen et al.(2001)]{Chen} 
Chen, P. S., Szczerba, R., Kwok, S., \& Volk, K. 2001, \aap, 368, 1006
\bibitem[Cioni (2009)]{Cioni}
Cioni, M.-R. L. 2009, \aap, 506, 1137
\bibitem[Claussen et al.(1987)]{Claussen}
Claussen, M. J., Kleinmann, S. G., Joyce, R. R., \& Jura, M. 1987, \apjs, 65, 385
\bibitem[Cole \& Weinberg(2002)]{Cole}
Cole A. A., \& Weinberg, M. D. 2002, \apj, 574, 43
\bibitem[Draine et al.(2007)]{PAHdust}
Draine, B. T., Dale, D. A., Bendo, G., et al. 2007, \apj, 663, 866
\bibitem[Egan et al.(2001)]{Egan}
Egan, M. P., Van Dyk, S. D., \& Price, S. D. 2001, \aj, 122, 1844
\bibitem[Genhard(2002)]{Gerhard}
Gerhard, O. 2002, The Dynamics, Structure \& History of Galaxies (ASP Conf. Ser. 273), ed. G. S. Da Costa \& H. Jerjen (San Francisco, CA: ASP), 73
\bibitem[Guillermo et al.(2001)]{GHZ} 
Guillermo, G., Donald, B., \& Peter, W. 2001, Icarus, 152, 185
%
\bibitem[Habing et al.(1985)]{ARAA}
Habing, H. J., Olnon, F. M., Chester, T., Gillett, F., \& Rowan-Robinson, M.
1985, \aap, 152, 1
\bibitem[Ishihara et al.(2006)]{ScanOpe}
Ishihara, D., Wada, T., Onaka, T., et al. 2006, \pasp, 118, 324
\bibitem[Ishihara et al.(2010)]{MirCat}
Ishihara, D., Onaka, T., Kataza, H., et al. 2010, \aap, 59, S443
%
\bibitem[Ishihara et al.(2010)]{Tycho}
Ishihara, D., Kaneda, H., Furuzawa, A., et al. 2010, \aap, 521, 61
%
\bibitem[Ita et al.(2010)]{Yita}
Ita, Y., Matsuura, M., Ishihara, D., et al. 2010, \aap, 514, 2
\bibitem[Jura \& Kleinmann(1990)]{Jura}
Jura, M., \& Kleinmann, S. G. 1990, \apj, 364, 663
\bibitem[Le Bertre et al.(2003)]{IRTS2}
Le Bertre, T., Tanaka, M., Yamamura, I., \& Murakami, H. 2003, \aap, 403, 943
\bibitem[Lee et al.(2009)]{HoGyu} 
Lee, H. G., Koo, B. C., Moon, D. S., et al. 2009, \apj, 706, 441
\bibitem[Lumsden et al.(2002)]{Lumsden}
Lumsden, S. L., Hoare, M. G., Oudmaijer, R. D., \& Richards, D. 2002, \mnras, 336, 621
\bibitem[Matsuura et al.(2009)]{Mikako}	
Matsuura, M., Barlow, M. J., Zijlstra, A. A., et al. 2009, \mnras, 396, 918
\bibitem[Murakami et al.(1996)]{IRTS}
Murakami, H., Freund, M. M., Ganga, K., et al. 1996, \pasj, 48, L41
\bibitem[Murakami et al.(2007)]{AKARI}
Murakami, H., Baba, H., Barthel, P., et al. 2007, \pasj, 59, 369
\bibitem[Neugebauer \& Leighton(1969)]{TMSS}
Neugebauer, G., \& Leighton, R. B. 1969, {\em Two Micron Sky Survey}
(NASA SP-2047)(TMSS)
\bibitem[Neugebauer et al.(1984)]{IRAS}
Neugebauer, G., Habing, H. J., van Duinen, R., et al. 1984, \apj, 278, 1
\bibitem[Noda et al.(1996)]{NIRS}
Noda, M., Matsumoto, T., Murakami, H., et al. 1996, \procspie, 2817, 248
\bibitem[Noguchi et al.(2004)]{Noguchi}
Noguchi, K., Aoki, W., \& Kawanomoto, S. 2004, \aap, 418, 67
\bibitem[Onaka et al.(1996)]{OnakaIRTS}
Onaka, T., Yamamura, I., Tanabe, T., Roellig, T. L., \& Yuen, L. 1996, \pasj, 48, 59
\bibitem[Onaka et al.(2007)]{IRC}
Onaka, T., Matsuhara, H., Wada, T., et al. 2007, \pasj, 59, S401
\bibitem[Pedicelli et al.(2009)]{Pedicelli}
Pedicelli, S., Bono, G., Lemasle, B., et al. 2009, \aap, 504, 81
\bibitem[Pritchet et al.(1987)]{Pritchet}
Pritchet, C. J., Schade, D., Richer, H. B., Crabtree, D., \& Yee, H. K. C. 1987, \apj, 323 79
\bibitem[Perryman et al.(1997)]{Hip}
Perryman, M. A. C., Lindegren, L., Kovalevsky, J., et al. 1997, \aap, 323, 49
\bibitem[Price et al.(2001)]{MSX}
Price, S. D., Egan, M. P., Carey, S. J., et al. 2001, \aj, 121, 2819
\bibitem[Rho et al.(2008)]{CasA} 
Rho, J., Kozasa, T., Reach, W.R., et al. 2008, \apj, 673, 271
\bibitem[Rudolph et al.(2006)]{Rudolph} 
Rudolph, A. L., Fich, M. B., Gwendolyn R., et al. 2006, \apjs, 162, 346
\bibitem[Shaver et al.(1983)]{Shaver}
Shaver, P. A., McGee, R. X., Newton, L. M., Danks, A. C., \& Pottasch, S. R.
1983, \mnras, 204, 53
\bibitem[Skrutskie et al.(2006)]{2MASS}
Skrutskie, M. F., Cutri, R. M., Stiening, R., et al. 2006, 131, 1163
\bibitem[Sloan et al.(2003)]{Sloan}
Sloan, G. C., Kraemer, K. E., Price, S. D., \& Shipman, R. F. 2003, \apjs, 147, 379
\bibitem[Smartt \& Rolleston(1997)]{Smartt} 
Smartt, S. J., \& Rolleston, W. R. J. 1997, \apj, 481, 47
\bibitem[Takita et al.(2010)]{Takita}
Takita, S., Kataza, H., Kitamura, Y., et al. 2010, \aap, 519, 83
\bibitem[Taylor \& Cordes(1993)]{Arms}
Taylor, J. H., \& Cordes, J. M. 1993, ApJ, 411, 674
\bibitem[Thronson et al.(1987)]{Thronson}
Thronson, H. A., Jr., Latter, W. B., Black, J. H., Bally, J., \&
Hacking, P. 1987, \apj, 322, 770
\bibitem[van der Veen \& Habing(1988)]{IRAScc}
van der Veen, W. E. C. J., \& Habing, H. J. 1988, \aap, 194, 125
1992, \apjs, 83, 111
\bibitem[Walker \& Cohen(1988)]{WalkerA} 
Walker, H. J., \& Cohen, M. 1988, \aj, 95, 1801
\bibitem[Walker et al.(1989)]{WalkerB} 
Walker, H. J., Volk, K., Wainscoat, R. J., et al. 1989, \aj, 98, 2163
\bibitem[Weingartner \& Draine(2001)]{Av}
Weingartner, J. C., \& Draine, B. T. 2001, \apj, 548, 296
\bibitem[Westerlund(1965)]{Westerlund}
Westerlund, B. E. 1965, \mnras, 130, 45
\bibitem[Wright et al.(2010)]{WISE}
Wright, E. L., Eisenhardt, P. R. M., Mainzer, A. K., et al. 2010, \aj, 140, 1868
\bibitem[Zhang et al.(2010)]{Zhang}
Zhang, H., Zhou J., Dong, G., Esimbek, J., \& Mu, J. 2010, \apss, 330, 23
\end{thebibliography}
\end{document}